%% file: main_jbr_twocolumns.tex
\documentclass[compact,twocolumn]{jetbrains_report}

\usepackage[style=authoryear,backend=biber,natbib=true]{biblatex}
\IfFileExists{bibliography.bib}{\addbibresource{bibliography.bib}}{}
\IfFileExists{yaroslav_zotero.bib}{\addbibresource{yaroslav_zotero.bib}}{}
\usepackage[capitalize,noabbrev]{cleveref}

\tcbuselibrary{listings, breakable}
\newtcblisting{promptbox}{
  breakable, fontupper=\footnotesize\ttfamily,
  listing only, colback=gray!8, colframe=gray!50,
}
\title{Towards Evaluation of Implicit Software World Models in Coding LLMs}

\jbauthor{Egor Bogomolov}{1,*}
\jbauthor{Yaroslav Zharov}{1,*}
\jbaffiliation{1}{JetBrains Research}
\jbcorresponding*[egor.bogomolov@jetbrains.com,\;yaroslav.zharov@jetbrains.com]

\jbmetadata{Data}{\href{https://huggingface.co/collections/JetBrains-Research/dl4c26-evaluation-of-software-world-models}{https://huggingface.co/collections/JetBrains-Research/dl4c26-evaluation-of-software-world-models}}
\jbmetadata{Code}{\href{https://github.com/JetBrains-Research/cwm-execution-tracer}{https://github.com/JetBrains-Research/cwm-execution-tracer}}
\jbnotice{\textsuperscript{*}Equal contribution; author order was determined by a coin toss.}

\begin{document}

\maketitle

\begin{jbabstract}
Software engineering, whether performed by humans or by AI agents, requires reasoning about how software behaves.
We call the internal model that supports such reasoning the \emph{software world model}, and view current code-execution benchmarks as covering one well-studied slice of it---control flow.
In this paper, we take a step toward a broader evaluation by shifting the observable axis to execution resources: alongside test outcome and exception class, we predict peak memory, wall-clock time, and ranked profiler outputs at method and line granularity.
We use SWE-bench Verified as the source of data to hold the test close to real-world software engineering tasks.
All tested models, frontier ones included, show modest performance and brittle behaviour, suggesting a notable lack of understanding of how software is \emph{executed}, as opposed to how its source code is \emph{written}.
\end{jbabstract}

\section{Introduction}
\label{sec:intro}

The most straightforward way of assessing coding LLMs is by how well they write code.
Function-level benchmarks such as HumanEval~\citep{humaneval2021} and MBPP~\citep{mbpp2021} score whether a generated solution passes a held-out test suite;
repository-level benchmarks such as SWE-Bench~\citep{swebench2024} and Aider-Polyglot~\citep{aiderpolyglot} score whether a generated patch fixes a real bug or implements a feature;
and repository-scale benchmarks such as Commit0~\citep{commit0_2024} ask a model to reproduce an entire Python repository from its specification and unit-test suite.
These evaluations measure the capabilities of model in code generation.

Another line of benchmarks evaluates coding LLMs on their understanding of program execution.
CRUXEval~\citep{cruxeval2024} predicts function inputs given outputs (and vice versa) on 800 short Python functions.
REval~\citep{reval2025} decomposes function execution into predicting code coverage, program state along the path, the next executed statement, and the final output.
ThrowBench~\citep{throwbench2025} predicts the runtime exception type.
BigO(Bench)~\citep{bigobench2025} classifies asymptotic complexity.
CodeMind~\citep{codemind2025} bundles three reasoning tasks on $1{,}450$ short programs.
RE2-Bench~\citep{liu2025evaluating} evaluates reasoning code on repository-scale.
The reasoning benchmarks mostly operate on isolated, often synthetically created, Python functions, classes, or short programs, and mostly focus on return value or a derivative.

Reasoning about code execution in terms of the control flow is, however, only one slice of the broader skill software engineers exercise daily --- predicting how a build will resolve dependencies, how a test suite will behave under a given change, how a service will respond at runtime, how concurrent code will interleave, how a patch will interact with the surrounding repository.
We use the term \emph{software world model} for this broader internal model of software-system behaviour, of which the control flow is the most studied facet.
Following the established usage, we distinguish \emph{implicit world models}---the capability spontaneously acquired by coding LLMs trained on general code corpora---from \emph{explicit world models}, models trained for this purpose specifically, such as the recent CWM~\citep{cwm2025}.

A complete evaluation of an implicit software world model would span build and dependency resolution, test and CI behaviour, runtime errors, deployment and runtime environments, concurrency, and the agentic, repo-level workflows that connect them --- work well beyond a single paper.
As a first step in this direction, we stay within the most-studied facet, code execution, and push it toward more realistic software contexts.
We address two gaps in current code-execution evaluation.
First, the function-scoped snippets used in most reasoning benchmarks do not match the complexity of practical software, where behaviour depends on a surrounding library implementation.
Second, the return value is only one of many facets of how a piece of software actually executes.
To bridge these gaps, we (1) design a set of metrics that captures a wider slice of execution behaviour, (2) collect a dataset of library-level cases derived from SWE-bench Verified, and (3) present results for a broad set of recent code-fluent LLMs, both proprietary and open-weight.
We position this contribution as a template for further extensions of software world model evaluation, beyond code execution alone. The data package is available on Hugging Face\footnote{\url{https://huggingface.co/collections/JetBrains-Research/dl4c26-evaluation-of-software-world-model}}, and the code on GitHub\footnote{\url{https://github.com/JetBrains-Research/cwm-execution-tracer}}.

\section{Data}
\label{sec:data}

We build the dataset from SWE-bench Verified~\citep{swebench2024}, a curated collection of 500 real GitHub issues and verified gold patches across 12 Python repositories.
We build samples from the pairs of tests that were failing and then got fixed with the gold patch.
To make the task solvable for most models, we only retain the examples that require less than 500K characters of context.
As SWE-bench Verified is dominated by samples from Django, we then downsample the data to 435 examples, while preserving diversity at the repository level.
To collect the ground truth observables, for each instance we inject a custom \texttt{sys.settrace}/\texttt{sys.monitoring}-based tracer into the SWE-bench Docker container, and run the designated tests.
We run the procedure before and after the patch, to catch different behaviours of the same test.
The tracer records a number of observables detailed further in this section.

\textbf{Test outcome.}
The tracer records whether each test passed or failed.
We count a test as failed if it raises an \texttt{AssertionError} or another exception.
For tests that failed, we additionally record the exception class name.

\textbf{Wall-clock time.}
We record the wall-clock time as the elapsed time between the start and finish of the function execution.
To alleviate the noise from other measurements, we do this on a clean run without any additional tracing.

\textbf{Peak memory.}
Memory required for the test to run.
We record it as a maximum of memory consumption recorded during the test execution, measured at the line level.

\textbf{Profiler.}
For each run, we record four profiling types, formed by the Cartesian product of two profiling scopes (method-level and line-level) and two profiling metrics (time and memory).
We record wall-clock time in milliseconds and memory in kilobytes.
For the method profiler, we record fully qualified function names, and for the line profiler, we record lines in the format \texttt{<file>:<line>}.
In all cases, we record the top 20 rows, ordered by the respective metric, and limit the scope of profiling to the repository.

\section{Metrics}
\label{sec:metrics}

We target three qualitatively different prediction tasks and use a distinct metric family for each.
Outcome prediction is evaluated by classification accuracy; resource prediction by linear calibration on a log scale, since wall time and memory span several orders of magnitude; and profiling as a ranking task.
All metrics are computed per instance and aggregated over the dataset.

\textbf{Test failure.}
We treat test failure prediction as a binary classification and report precision, recall, and F1.

\textbf{Linear calibration for wall time and peak memory.}
Both quantities span several orders of magnitude, so we evaluate on the $\log_{10}$ scale.
We fit a linear model $\hat{y} = a\,y^{*} + b$ where $\hat{y} = \log_{10}(\text{predicted})$ and $y^{*} = \log_{10}(\text{actual})$, and report slope $a$ (calibration), intercept $b$ (systematic bias), and Mean Absolute Error.
We additionally replace zero predictions of models with 0.01 ms for time and 10KB for peak memory.
These values are smaller than all the groundtruth numbers in our dataset, yet they allow us to compute log-scale metrics.

\textbf{Profiler ranking.}
For each of the four ranked lists (functions $\times$ \{time, memory\} and lines $\times$ \{time, memory\}) we report two ranking metrics: recall and NDCG.
$\mathrm{Recall}@k$ yields $1$ for a sample if the actual top-method/-line is within top $k$ predicted methods/lines and $0$ otherwise.
$\mathrm{DCG}@k$ is defined as $\sum_{i=1}^{k} r_i / \log_2(i+1)$, where $r_i$ is the measured time/memory attributed to the function or line predicted at rank $i$.
We report $\mathrm{NDCG}@5$---$\mathrm{DCG}@5$ normalized by the optimal $\mathrm{DCG}@5$---and $\mathrm{Recall}@5$.
To gauge the model's ability to estimate the execution scope we additionally measure execution rate, which defines a portion of predicted methods or lines that were actually executed during the run.

\section{Experiment Setup}
\label{sec:experiment-setup}

\textbf{Models.}
We evaluate three Anthropic models via the Anthropic API: \texttt{claude-haiku-4-5}~\citep{claudehaiku45card2025}, \texttt{claude-sonnet-4-6}~\citep{claudesonnet46card2026}, and \texttt{claude-opus-4-7}~\citep{claudeopus47card2026}.
Via the OpenAI API we evaluate \texttt{gpt-5-mini}~\citep{gpt5card2025}, \texttt{gpt-5.2}~\citep{gpt52card2025}, \texttt{gpt-5.4}~\citep{gpt54card2026}, and \texttt{gpt-5.5}~\citep{gpt55card2026}.
We run five open-weight models locally: \texttt{gpt-oss-120b}~\citep{gptosscard2025} from OpenAI; \texttt{Qwen3.5-397B-A17B}~\citep{qwen35blog2026}, \texttt{Qwen3-235B-A22B-Instruct}, and \texttt{Qwen3-30B-A3B-Instruct}~\citep{qwen3blog2025} from Alibaba; and \texttt{CWM}~\citep{cwm2025} by FAIR.

\textbf{Context.}
The user message consists of two blocks.
The \textbf{first block} is a slice of the library containing all executed code. For each file, where at least one line was executed, we include module preambles (all content before the first class or function definition), followed by every function (or method body enclosed in class scaffolding) in which at least one line was executed.
For classes with executed lines we also keep everything outside methods such as class definition and its fields.
The remaining budget (if any) up to 500\,K characters is then filled with non-executed functions and methods.
The \textbf{second block} is the test file, windowed to at most 60\,K characters centered on the target test function.
The prompt template is shared in~\Cref{app:prompt}.

\textbf{Task.}
The model is instructed to return a single JSON object with ten fields: \textit{reasoning} (a 2--4 sentence explanation), \textit{outcome} (\textit{passed}, \textit{failed}, or \textit{error}), \textit{failure\_line} and \textit{exception\_type} (null when outcome is passed), \textit{peak\_bytes} and \textit{wall\_ms} (integer and float respectively), and four ranked lists of up to 20 entries each: \textit{hot\_methods\_time}, \textit{hot\_methods\_alloc}, \textit{hot\_lines\_time}, and \textit{hot\_lines\_alloc}.
These fields correspond directly to the observables described in Section~\ref{sec:data} and are scored by the metrics in Section~\ref{sec:metrics}.

\section{Results Discussion}
\label{sec:results}

\input{reports_tables_yaroslav}

\Cref{tab:outcome} reports test-outcome classification; \Cref{tab:peak-rss,tab:wall-time} report resource-prediction calibration; \Cref{tab:mem-profiler,tab:time-profiler} report profiler-ranking quality.
We note that all the models, including the frontier models available by API aren't scoring high on the proposed tasks.
For test outcome prediction, most models have F1 scores consistently low, largely due to very low recall.
For both peak memory and wall-time, model predictions are systematically biased, with every model having slope and bias significantly different from optimal $(1,0)$ values.
For all types of profiling, the best $recall@5$ never reaches $0.2$, indicating that models rarely identify the most consuming entity.

For the test outcome prediction, we note that most models have low recall, indicating a strong bias towards tests passing.
We attribute this to LLMs' tendency to follow the natural-language semantics of code rather than its structure, as shown by~\citet{lam_codecrash_2025}.
Optimizing tests to elicit unbiased predictions from LLMs may be a promising research direction for Software Engineering.

For the peak memory consumption and time, in addition to the presence and strength of systematic errors, we note that the bias itself is universal.
We observe unanimous slope compression (models giving predictions closer to the average) and bias towards overestimation (for most of the models the difference is at least an order of magnitude).
The unique case of CWM can be explained by 247 cases where it predicts 0 ms execution time, motivating it in reasoning by inability to calculate it precisely.
This conservative estimations make models less useful in giving predictions when it comes to load brought by particular tasks\footnote{On an anecdotal observation, while we were running experiments for this paper, the runs were supervised by Sonnet 4.6, which systematically mispredicted the time needed to complete a run by always predicting a value around 20 minutes, whether in reality it took 2 hours or 7 minutes.}.

Most often, the predictions for the profiling tasks are dominated by the line numbers and methods that weren't actually executed.
We additionally measured NDCG@5 of a set obtained by correct ordering of model predictions and noticed a significant boost (On average $\times 1.5$ for method profiling and $\times 2$ for line profiling), which indicates that hallucinating execution scope is not the sole problem --- correct ranking is challenging for models.

In general, we note that smaller and especially open-weight models have a tendency to predict "round" numbers, with particularly peculiar numbers like "12345 ms" dominating the output of Qwen3-30B.
This is especially well notable on the scatter plots of predicted and measured values shown in~\Cref{app:scatter}.
Surprisingly, CWM's performance on code generation and reasoning, is not transferring to reasoning about software---while it produces coherent reasoning traces and outputs, it falls behind the Qwen3-30B (a model of the similar size) in most experiments.
These findings reinforce our point about the need for a wider understanding of Software World Models beyond Code World Models.

\section{Limitations}
\label{sec:limit}
We see three points where further work can make the results more diverse: data, context, and answer elicitation techniques.
The dataset is derived from SWE-bench Verified and further limits its scope by filtering the tests and libraries where the context doesn't fit in 500k characters.
We leave generalisation to other languages, less-curated codebases, or broader test populations to future work.
This paper only evaluates a single oracle-based context-collection strategy, thus establishing the upper boundary of the context collection performance.
Future work should explore more realistic scaffolding.
Finally, this paper does not explore broader ways to elicit better answers from LLMs, leaving such strategies as advanced prompting, multi-shot voting, or probing the latent space open for further exploration.

\section{Conclusion}
\label{sec:conc}

Code execution is one facet of the broader \emph{software world modeling}; this paper is intended as a first probe into the rest of that space.
We extend execution evaluation to library-level cases from SWE-bench Verified and to four tasks beyond the return value: test outcome, peak memory consumption, wall time, and ranked profiler outputs at method and line granularity.
Across twelve models, including the trace-trained \texttt{CWM} and frontier models, performance is modest.
We envision further extensions to the area of software world modeling in such tasks as build resolution, CI, deployment, concurrency, and agentic workflows; we release the data, prompts, and tracing harness to contribute to help advance the research in this area.

\printbibliography

\onecolumn
\appendix
\crefalias{section}{appendix}
\IfFileExists{appendix_prompt.tex}{\input{appendix_prompt}}{}
\IfFileExists{appendix_scatter_yaroslav.tex}{\input{appendix_scatter_yaroslav}}{}

\end{document}

%% file: reports_tables_yaroslav.tex

\begin{table}[t]
\caption{Test failure prediction, sorted by F1\,$\downarrow$.}
\label{tab:outcome}
\centering
\small
\setlength{\tabcolsep}{4pt}
\begin{tabular}{lccc}
\toprule
Model & Prec.\,$\uparrow$ & Rec.\,$\uparrow$ & F1\,$\uparrow$ \\
\midrule
\texttt{gpt-5.5}             & \textbf{0.987} & \textbf{0.735} & \textbf{0.842} \\
\texttt{gpt-oss-120b}        & 0.773 & \underline{0.495} & \underline{0.604} \\
\texttt{claude-sonnet-4-6}   & 0.907 & 0.390 & 0.545 \\
\texttt{gpt-5-mini}          & 0.669 & 0.395 & 0.497 \\
\texttt{claude-opus-4-7}     & 0.821 & 0.345 & 0.486 \\
\texttt{Qwen3.5-397B}        & 0.762 & 0.320 & 0.451 \\
\texttt{gpt-5.2}             & 0.720 & 0.270 & 0.393 \\
\texttt{gpt-5.4}             & \underline{0.959} & 0.235 & 0.378 \\
\texttt{CWM}                 & 0.636 & 0.210 & 0.316 \\
\texttt{claude-haiku-4-5}    & 0.597 & 0.185 & 0.282 \\
\texttt{Qwen3-235B}          & 0.800 & 0.080 & 0.145 \\
\texttt{Qwen3-30B}           & 0.833 & 0.025 & 0.049 \\
\bottomrule
\end{tabular}
\end{table}

\begin{table}[t]
\caption{Peak memory consumption prediction calibration ($\log_{10}$ scale). Ideal: slope\,$=1$, bias\,$=0$. Sorted by MAE\,$\downarrow$.}
\label{tab:peak-rss}
\centering
\small
\setlength{\tabcolsep}{4pt}
\begin{tabular}{lccc}
\toprule
Model & Slope\,$\uparrow$ & Bias\,$\downarrow$ & MAE\,$\downarrow$ \\
\midrule
\texttt{gpt-5.4}             & 0.738 & \textbf{$+$1.176} & \textbf{0.567} \\
\texttt{gpt-5.2}             & 0.661 & $+$1.883 & \underline{0.631} \\
\texttt{Qwen3-30B}           & 0.312 & $+$2.844 & 0.723 \\
\texttt{gpt-5.5}             & \textbf{0.802} & \underline{$+$1.616} & 0.729 \\
\texttt{claude-sonnet-4-6}   & \underline{0.760} & $+$1.711 & 0.756 \\
\texttt{gpt-oss-120b}        & 0.733 & $+$1.765 & 0.789 \\
\texttt{Qwen3.5-397B}        & 0.660 & $+$1.920 & 0.828 \\
\texttt{claude-opus-4-7}     & 0.713 & $+$2.097 & 0.833 \\
\texttt{Qwen3-235B}          & 0.665 & $+$2.235 & 0.869 \\
\texttt{claude-haiku-4-5}    & 0.640 & $+$2.469 & 0.911 \\
\texttt{CWM}                 & 0.186 & $+$3.761 & 1.004 \\
\texttt{gpt-5-mini}          & 0.405 & $+$4.063 & 1.188 \\
\bottomrule
\end{tabular}
\end{table}

\begin{table}[t]
\caption{Wall-time prediction calibration ($\log_{10}$ scale). Ideal: slope\,$=1$, bias\,$=0$. Sorted by MAE\,$\downarrow$.}
\label{tab:wall-time}
\centering
\small
\setlength{\tabcolsep}{4pt}
\begin{tabular}{lccc}
\toprule
Model & Slope\,$\uparrow$ & Bias\,$\downarrow$ & MAE\,$\downarrow$ \\
\midrule
\texttt{gpt-5.4}             & 0.800 & \textbf{$+$0.500} & \textbf{0.578} \\
\texttt{gpt-5.5}             & \textbf{0.895} & \underline{$+$0.652} & \underline{0.686} \\
\texttt{gpt-5.2}             & 0.763 & $+$0.880 & 0.893 \\
\texttt{gpt-oss-120b}        & 0.630 & $+$0.917 & 0.944 \\
\texttt{Qwen3-235B}          & 0.405 & $+$1.057 & 1.021 \\
\texttt{Qwen3.5-397B}        & 0.626 & $+$1.154 & 1.078 \\
\texttt{gpt-5-mini}          & 0.442 & $+$1.148 & 1.095 \\
\texttt{claude-opus-4-7}     & 0.779 & $+$1.221 & 1.182 \\
\texttt{Qwen3-30B}           & 0.305 & $+$1.328 & 1.295 \\
\texttt{claude-sonnet-4-6}   & \underline{0.848} & $+$1.382 & 1.350 \\
\texttt{claude-haiku-4-5}    & 0.663 & $+$1.649 & 1.625 \\
\texttt{CWM}                 & 0.244 & $-$0.640 & 1.816 \\
\bottomrule
\end{tabular}
\end{table}

\begin{table*}[t]
\caption{Memory-profiler ranking quality. \emph{exec}: fraction of predicted names present in the execution trace.
NDCG@5 and recall@5 assess ranking quality against ground-truth allocation profiles.
Sorted by Method NDCG@5\,$\downarrow$.}
\label{tab:mem-profiler}
\centering
\small
\setlength{\tabcolsep}{5pt}
\begin{tabular}{lcccccc}
\toprule
& \multicolumn{3}{c}{\textbf{Method}} & \multicolumn{3}{c}{\textbf{Line}} \\
\cmidrule(lr){2-4}\cmidrule(lr){5-7}
Model & exec\,$\uparrow$ & NDCG@5\,$\uparrow$ & rec@5\,$\uparrow$ & exec\,$\uparrow$ & NDCG@5\,$\uparrow$ & rec@5\,$\uparrow$ \\
\midrule
\texttt{gpt-5.5}             & 0.222 & \textbf{0.127} & 0.136 & \textbf{0.349} & \textbf{0.095} & \textbf{0.094} \\
\texttt{gpt-5-mini}          & 0.254 & \underline{0.123} & \textbf{0.145} & 0.221 & \underline{0.057} & \underline{0.046} \\
\texttt{gpt-5.4}             & \underline{0.259} & 0.120 & \underline{0.141} & 0.169 & 0.017 & 0.011 \\
\texttt{gpt-5.2}             & \textbf{0.268} & 0.119 & 0.136 & 0.165 & 0.018 & 0.016 \\
\texttt{Qwen3.5-397B}        & 0.200 & 0.111 & 0.118 & 0.163 & 0.023 & 0.014 \\
\texttt{claude-opus-4-7}     & 0.219 & 0.108 & 0.108 & \underline{0.252} & 0.022 & 0.014 \\
\texttt{gpt-oss-120b}        & 0.199 & 0.103 & 0.111 & 0.138 & 0.020 & 0.023 \\
\texttt{claude-sonnet-4-6}   & 0.188 & 0.092 & 0.101 & 0.196 & 0.017 & 0.016 \\
\texttt{Qwen3-30B}           & 0.168 & 0.091 & 0.097 & 0.103 & 0.010 & 0.009 \\
\texttt{Qwen3-235B}          & 0.188 & 0.091 & 0.088 & 0.135 & 0.009 & 0.005 \\
\texttt{claude-haiku-4-5}    & 0.108 & 0.067 & 0.069 & 0.145 & 0.010 & 0.007 \\
\texttt{CWM}                 & 0.113 & 0.054 & 0.048 & 0.043 & 0.004 & 0.005 \\
\bottomrule
\end{tabular}
\end{table*}

\begin{table*}[t]
\caption{Time-profiler ranking quality. \emph{exec}: fraction of predicted names present in the execution trace.
NDCG@5 and recall@5 assess ranking quality against ground-truth time profiles.
Sorted by Method NDCG@5\,$\downarrow$.}
\label{tab:time-profiler}
\centering
\small
\setlength{\tabcolsep}{5pt}
\begin{tabular}{lcccccc}
\toprule
& \multicolumn{3}{c}{\textbf{Method}} & \multicolumn{3}{c}{\textbf{Line}} \\
\cmidrule(lr){2-4}\cmidrule(lr){5-7}
Model & exec\,$\uparrow$ & NDCG@5\,$\uparrow$ & rec@5\,$\uparrow$ & exec\,$\uparrow$ & NDCG@5\,$\uparrow$ & rec@5\,$\uparrow$ \\
\midrule
\texttt{gpt-5.4}             & \underline{0.309} & \textbf{0.197} & 0.154 & 0.178 & 0.021 & 0.007 \\
\texttt{gpt-5.5}             & 0.268 & \underline{0.188} & \textbf{0.168} & \textbf{0.361} & \textbf{0.111} & \textbf{0.106} \\
\texttt{gpt-5.2}             & \textbf{0.328} & 0.186 & 0.150 & 0.167 & 0.017 & 0.007 \\
\texttt{claude-opus-4-7}     & 0.294 & 0.184 & \underline{0.159} & \underline{0.260} & 0.039 & 0.023 \\
\texttt{gpt-5-mini}          & 0.305 & 0.183 & 0.154 & 0.209 & \underline{0.080} & \underline{0.085} \\
\texttt{Qwen3.5-397B}        & 0.255 & 0.177 & 0.145 & 0.165 & 0.024 & 0.018 \\
\texttt{gpt-oss-120b}        & 0.245 & 0.161 & 0.129 & 0.143 & 0.026 & 0.023 \\
\texttt{claude-sonnet-4-6}   & 0.253 & 0.159 & 0.136 & 0.200 & 0.027 & 0.016 \\
\texttt{Qwen3-30B}           & 0.219 & 0.153 & 0.122 & 0.106 & 0.012 & 0.005 \\
\texttt{Qwen3-235B}          & 0.241 & 0.152 & 0.138 & 0.139 & 0.014 & 0.007 \\
\texttt{claude-haiku-4-5}    & 0.152 & 0.130 & 0.127 & 0.147 & 0.018 & 0.007 \\
\texttt{CWM}                 & 0.138 & 0.077 & 0.067 & 0.048 & 0.004 & 0.000 \\
\bottomrule
\end{tabular}
\end{table*}

%% file: appendix_prompt.tex

\section{Prompt Template}
\label{app:prompt}

Each sample is sent to the model as a two-message conversation.
The system message is identical for all samples; the user message
is constructed per sample as described in \Cref{sec:experiment-setup}.
Variable parts are shown in \textlangle\textit{angle brackets}\textrangle.

\subsection*{System message}

\begin{promptbox}
You are an expert Python developer analyzing a software project. You will be shown a slice of the project's source code and the full test file. You will predict a specific runtime property of running that test against THIS source code.

You do NOT execute the code. Reason about it by reading the source. Respond with ONLY a JSON object — no prose, no markdown fences.
\end{promptbox}
\normalsize

\subsection*{User message}

\small
\begin{promptbox}
## Source files (slice of the project)

### `<rel/path/to/file.py>`
```python
<executed functions and module preamble, then non-executed functions up to 400 000 characters total across all files>
```

### `<rel/path/to/another_file.py>`
...

## Test file: `<rel/path/to/test_file.py>`

<test file content, up to 60 000 characters, windowed around the target test function>

## Task

Predict the runtime behavior of running the test <test_id> against the source code shown above. Return a single JSON object with all of the following keys:

  reasoning           — 2-4 sentences explaining your overall analysis
  outcome             — "passed", "failed" (AssertionError), or "error" (non-assertion exception)
  failure_line        — 1-based line in test file <test_file> where the failure occurs; null if outcome == passed
  exception_type      — exception class name ("AssertionError", "TypeError", ...); null if outcome == passed
  peak_bytes          — peak memory the test needs above its baseline, in bytes (int). i.e. how much additional RAM the system must have free for the test to run correctly. Concretely: the high-water mark of memory usage during test execution MINUS the memory already in use when the test started; taken as the larger of two complementary measurements:
                          * Python-heap (tracemalloc) peak delta — catches lists / dicts / strings.
                          * Process-RSS peak delta — catches large numpy / C-extension buffers that bypass pymalloc.
                        Allocate-then-free patterns count their peak, NOT the cumulative bytes: a loop of 100 iterations each allocating + freeing 80 MB has peak_bytes ≈ 80 MB. Process-baseline (imports + state from prior tests in the session) is NOT counted.
  wall_ms             — total wall-clock time to run the test, in milliseconds (float). What a stopwatch would show from when the test framework (pytest / unittest) invokes the test method to when it returns — the test method body plus any setUp / fixtures / tearDown. Includes time spent in stdlib, numpy, database drivers, network I/O, etc. Does NOT include test-runner collection or reporting time outside the test invocation.
  hot_methods_time    — up to 20 fully-qualified function names from this project, ranked by total time spent executing them during the test. Uses EXCLUSIVE wall time: time in each function's own body, with time in nested in-project calls credited to the callee. Time in stdlib / numpy / third-party calls invoked from a method IS credited to it (we don't trace into those frames). Hottest first. Synthetic frames (<lambda>, <listcomp>, <dictcomp>, <genexpr>) are eligible.
  hot_methods_alloc   — up to 20 fully-qualified function names from this project, ranked by total bytes ALLOCATED (directly or indirectly via library calls) while executing during the test. Counts EVERY allocation event, including transient allocations that are freed before the function returns — a method that builds a 100 MB array, uses it, and discards it inside one call gets full credit. Combines Python-heap (tracemalloc) and process-RSS deltas to capture pymalloc AND C-extension buffers. Exclusive: allocations in stdlib / numpy / third-party calls invoked from this method ARE credited to it (not traced), but allocations inside in-project child methods are credited to those children. Hottest allocator first.
  hot_lines_time      — up to 20 <rel_file_path>:<line_number> strings, ranked by total wall time spent executing that line during the test (summed across every execution of the line). Use the paths from the source slice; only lines in those files are eligible. Hottest first.
  hot_lines_alloc     — up to 20 <rel_file_path>:<line_number> strings, ranked by total bytes allocated when that line executes (summed across every execution of the line). Captures the line's own allocation activity plus any library / stdlib allocations made by code called from that line. Largest allocator first.

Use "reasoning" to think before committing to a value. Return fewer than 20 entries in any list if you expect fewer than many to be relevant. Wrong names or paths in the lists score 0 for that slot.

Return JSON:
{
  "reasoning": "<2-4 sentences>",
  "outcome": "passed" | "failed" | "error",
  "failure_line": <int> | null,
  "exception_type": "<ClassName>" | null,
  "peak_bytes": <int>,
  "wall_ms": <float>,
  "hot_methods_time": ["fn1", "fn2", ...],
  "hot_methods_alloc": ["fn1", "fn2", ...],
  "hot_lines_time": ["path/to/file.py:42", ...],
  "hot_lines_alloc": ["path/to/file.py:42", ...]
}
\end{promptbox}
\normalsize

%% file: appendix_scatter_yaroslav.tex

\section{Calibration Scatter Plots}
\label{app:scatter}

\begin{figure*}[p]
\centering
\includegraphics[width=\textwidth]{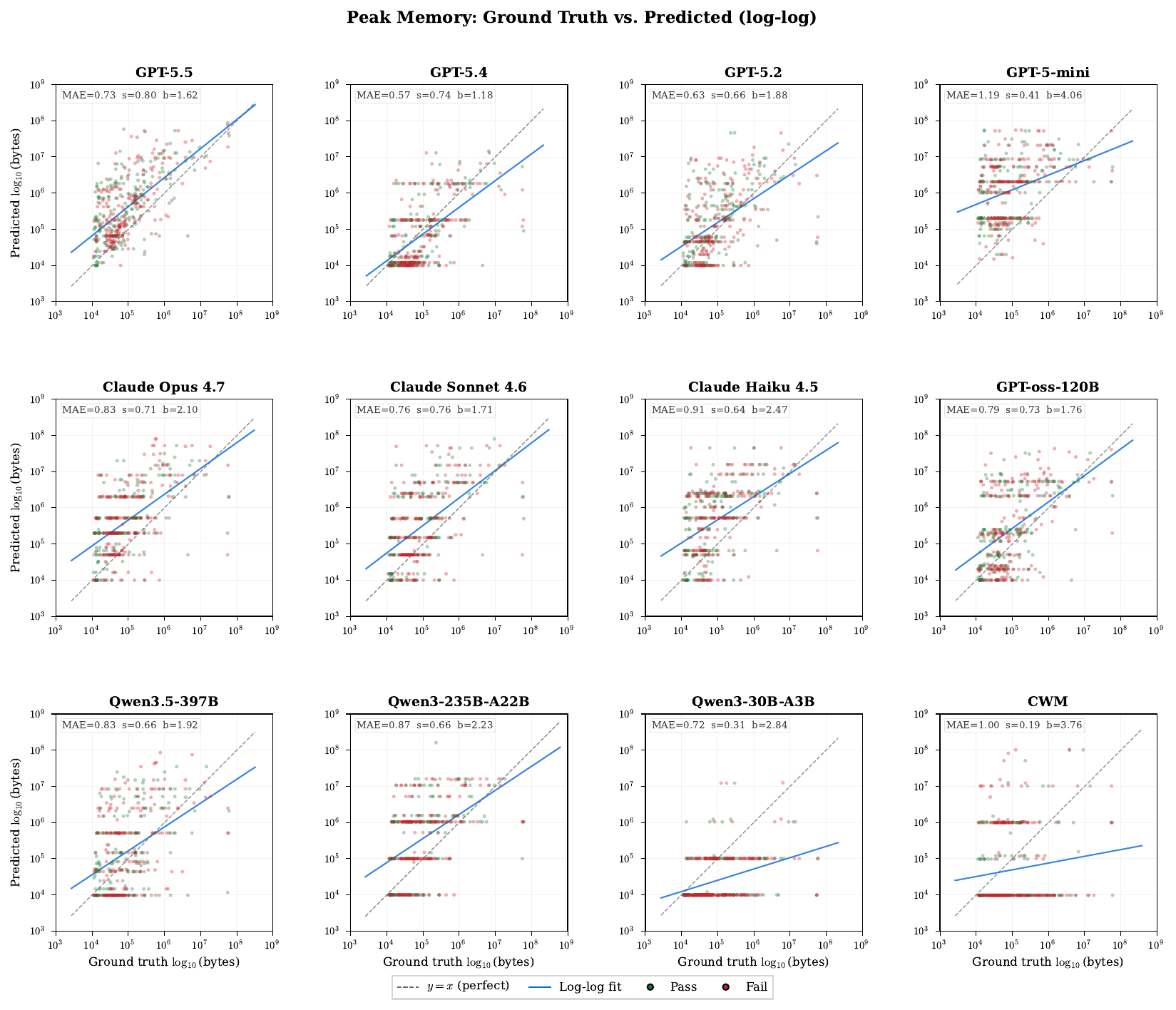}
\caption{%
  Peak heap allocation: predicted vs.\ ground-truth $\log_{10}(\text{bytes})$
  for all twelve models.
  Each panel's legend reports MAE, fitted slope $s$, and bias $b$;
  dashed diagonal is $y{=}x$ (perfect calibration).
  Points are coloured by test outcome (green = pass, red = fail).
}
\label{fig:scatter-peak-rss}
\end{figure*}

\begin{figure*}[p]
\centering
\includegraphics[width=\textwidth]{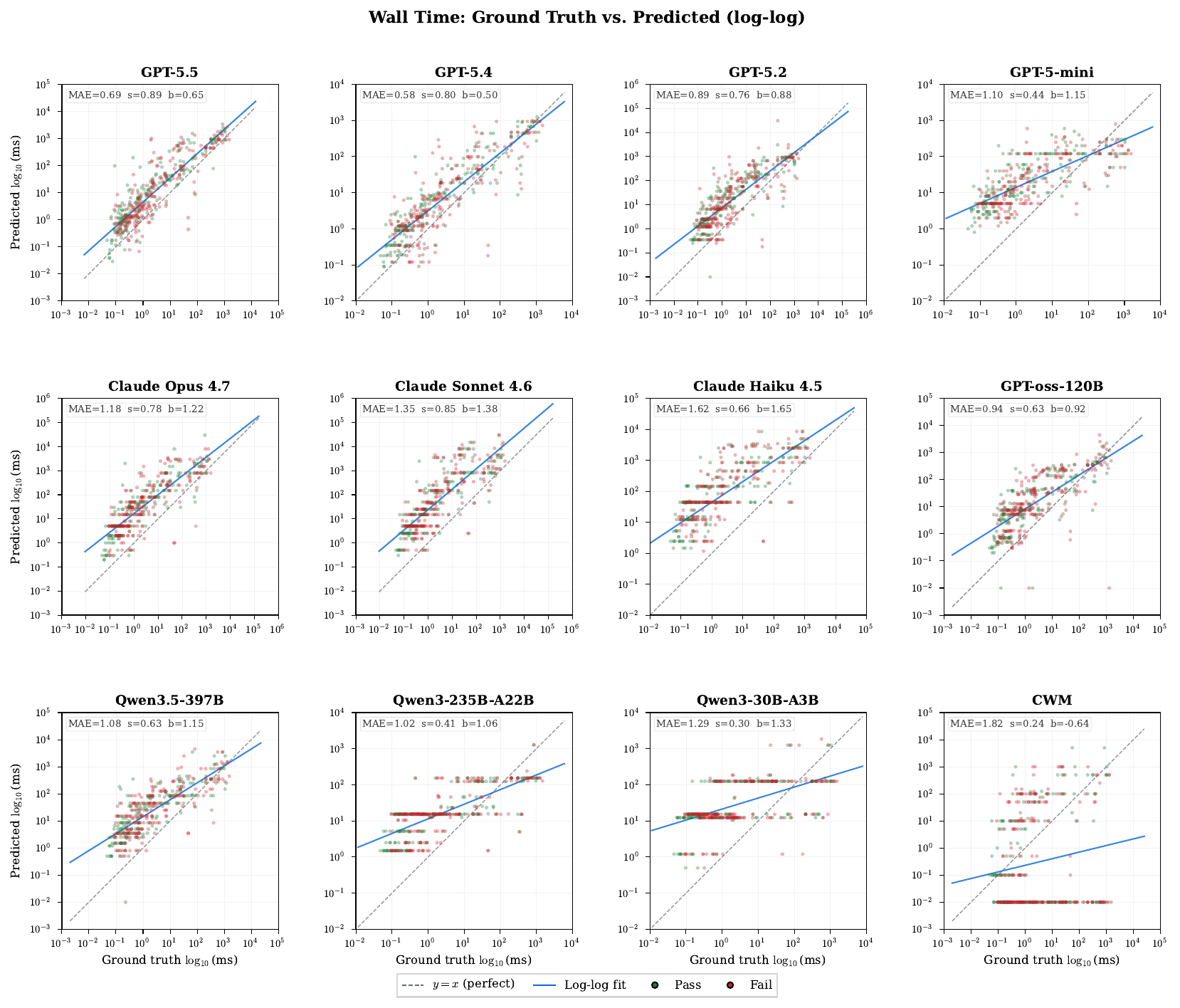}
\caption{%
  Wall-clock time: predicted vs.\ ground-truth $\log_{10}(\text{ms})$
  for all twelve models.
  Layout and colour coding identical to Figure~\ref{fig:scatter-peak-rss}.
  CWM is the only model whose bias is negative ($b{=}{-}0.64$),
  reflecting systematic under-prediction of execution time.
}
\label{fig:scatter-wall-time}
\end{figure*}

Figures~\ref{fig:scatter-peak-rss} and~\ref{fig:scatter-wall-time} show
per-model scatter plots of predicted versus ground-truth values for peak
heap allocation and wall-clock time, respectively.
Each panel reports the log-log linear fit
($\hat{y} = s \cdot x + b$, slope $s$ and bias $b$) together with the
mean absolute log\textsubscript{10} error (MAE).
The dashed diagonal marks perfect calibration ($y{=}x$); the solid blue
line is the fitted regression.
Points are coloured by ground-truth test outcome (pass / fail).
All models consistently overestimate both quantities (positive bias),
with slope below~1 indicating compression of the dynamic range.

